\documentclass[aps,pre,twocolumn]{revtex4}

\usepackage[english]{babel}
\usepackage[utf8x]{inputenc}
\usepackage{amsmath,amsfonts,amssymb,textcomp}
\usepackage{hyperref}
\usepackage{color}
\usepackage{array}
\usepackage{graphicx}
\usepackage{natbib}
\usepackage{rotating}

\graphicspath{{./figures/}}

\newcommand{\FIG}[1]{Fig.~\ref{fig:#1}}

\definecolor{drab}{rgb}{0.59, 0.44, 0.09}

\definecolor{celestialblue}{rgb}{0.29, 0.59, 0.82}

\definecolor{purple}{rgb}{0.459,0.109,0.538}
\definecolor{deepsaffron}{rgb}{1.0, 0.6, 0.2}

\begin{document}
\title{Establishment and stability of the latent HIV-1 DNA reservoir}

\author{Johanna Brodin$^{1}$, Fabio Zanini$^{2,3}$, Lina Thebo$^{1}$, Christa Lanz$^{2}$, G\"oran Bratt$^4$, Richard A.~Neher$^{2}$, Jan Albert$^{1,5}$}
\affiliation{$^{1}$Department of Microbiology, Tumor and Cell Biology, Karolinska Institute, Stockholm, Sweden\\
$^{2}$Max Planck Institute for Developmental Biology, 72076 T\"ubingen, Germany\\
$^{3}$current address: Stanford University, Stanford, CA, USA\\
$^4$Department of Clinical Science and Education, Venh\"alsan, Stockholm South General Hospital, Stockholm, Sweden\\
$^5$Department of Clinical Microbiology, Karolinska University Hospital, Stockholm, Sweden
}
\date{\today}

\begin{abstract}
HIV-1 infection currently cannot be cured because the virus persists as integrated proviral DNA in long-lived cells despite years of suppressive antiretroviral therapy (ART).
To characterize establishment, turnover, and evolution of viral DNA reservoirs we deep-sequenced the p17gag region of the HIV-1 genome from samples obtained from 10 patients after 3-18 years of suppressive ART.
For each of these patients, whole genome deep-sequencing data of HIV-1 RNA populations before onset of ART were available from 6-12 longitudinal plasma samples spanning 5-8 years of untreated infection. This enabled a detailed analysis of the dynamics and origin of proviral DNA during ART.
A median of 14\% (range 0-42\%) of the p17gag DNA sequences were overtly defective due to G-to-A hypermutation.
The remaining sequences were remarkably similar to previously observed RNA sequences and showed no evidence of evolution over many years of suppressive ART.
Most sequences from the DNA reservoirs were very similar to viruses actively replicating in plasma (RNA sequences) shortly before start of ART. The results do not support persistent HIV-1 replication as a mechanism to maintain the HIV-1 reservoir during suppressive therapy. Rather, the data indicate that viral DNA variants are turning over as long as patients are untreated and that suppressive ART halts this turnover.
\end{abstract}
\maketitle

\section*{Introduction}
Combination antiretroviral therapy (ART) has had a dramatic effect on the morbidity and mortality of human immunodeficiency virus type 1 (HIV-1) infection. Even though ART
is very effective in suppressing active virus replication, it cannot eradicate the infection because HIV-1 persists as integrated proviral DNA in long-lived cells that constitute a virus reservoir. Latently infected resting memory CD4+ T-lymphocytes (memory CD4 cells) represent the most solidly documented HIV-1 reservoir \cite{eriksson_comparative_2013,chun_quantification_1997,chun_vivo_1995}. Thus, a small fraction of memory CD4 cells have fully functional integrated HIV-1 proviruses. These cells do not produce virus when they are in a resting state, but can be induced to produce virus upon activation in vitro and in vivo \cite{eriksson_comparative_2013,chun_vivo_1995,chun_quantification_1997,massanella_measuring_2016}.

Because of their importance for HIV-1 cure efforts, many methods to quantify the HIV-1 reservoirs have been developed. The quantitative virus outgrowth assay (QVOA) represents the ``gold standard'' \cite{stockenstrom_longitudinal_2015,massanella_measuring_2016,bruner_towards_2015}, but this assay underestimates the true size of the functional reservoir due to incomplete induction by PHA stimulation. Ho et~al.\cite{ho_replication-competent_2013} showed that the functional HIV-1 reservoir may be 60-fold larger than originally estimated.
PCR based assays are also commonly used for quantifying the HIV-1 reservoir, but these assays overestimate the size of the functional reservoir because they cannot distinguish between replication-competent and defective viral genomes. Quantification of the HIV-1 reservoir by PCR-based methods typically give at least 100-fold higher numbers that the QVOA because of defective proviruses \cite{stockenstrom_longitudinal_2015,massanella_measuring_2016,bruner_towards_2015}. Many defective proviruses have large internal deletions \cite{ho_replication-competent_2013,sanchez_accumulation_1997}. Defective proviruses are also the result of APOBEC editing, which induces G-to-A hypermutation \cite{yu_single-strand_2004,kieffer_g-->hypermutation_2005,stopak_hiv-1_2003}.

The HIV-1 reservoir is established early during primary infection and is remarkably stable in both quantitative and qualitative terms. Early ART reduces the size and the genetic complexity of the reservoir \cite{chomont_hiv_2009,josefsson_hiv-1_2013,lori_treatment_1999,strain_effect_2005}. Siliciano et al.\cite{siliciano_long-term_2003} documented a half-life of 44 months for latently infected cells capable of producing replication-competent virus in the QVOA. Similarly, HIV-1 DNA levels and genetic compositions are very stable in patients on long-term suppressive ART \cite{stockenstrom_longitudinal_2015,besson_hiv-1_2014,josefsson_hiv-1_2013,kearney_lack_2014,gunthard_evolution_1999,evering_absence_2012,kieffer_absence_2004}. Most studies indicate that the HIV-1 reservoir is maintained by the physiological homeostasis of CD4 memory that in part involves occasional expansions and contractions of individual CD4 cell clones \cite{stockenstrom_longitudinal_2015,chomont_maintenance_2011,chomont_hiv_2009}. However, some studies have suggested that persistent virus replication may be an important contributor to the maintenance of the HIV-1 reservoir \cite{buzon_hiv-1_2010,yukl_effect_2010}. In particular, Lorenzo-Redondo et al.\cite{lorenzo-redondo_persistent_2016} recently reported evidence of rapid HIV-1 evolution in lymphoid tissue reservoirs.

\begin{figure}[tb]
    \centering
    \includegraphics[width=\columnwidth]{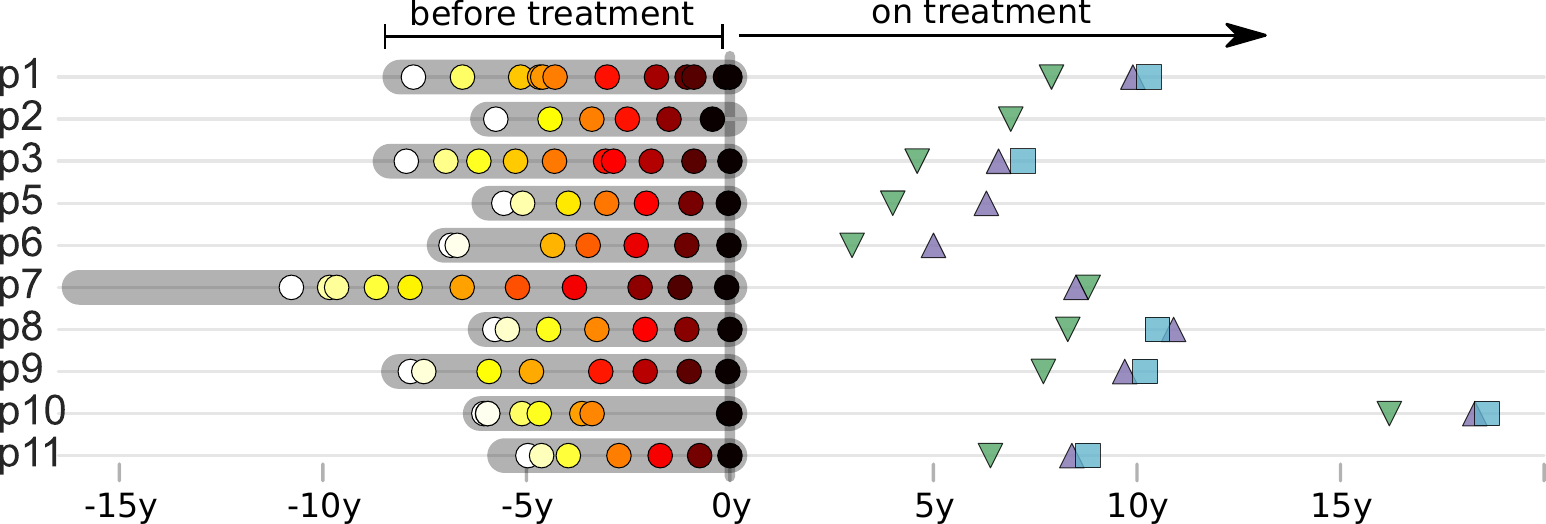}
    \caption{Sampling times before and after start of treatment. For each study participant, the thick grey bar indicates the period of untreated HIV-1 replication, while circles mark the collection times of RNA samples used for whole genome deep sequencing \cite{zanini_population_2016}. The collection dates of PBMC samples used for p17gag deep sequencing of integrated provirus are indicated by triangles and squares. All times are relative to start of treatment. }
    \label{fig:study_design}
\end{figure}

Despite their significance for HIV-1 cure efforts relatively little is known about the establishment and turnover of the HIV-1 reservoir before start of ART.
In this study we have characterized how HIV-1 DNA reservoirs are established and maintained in 10 patients. Evolution of HIV-1 in these patients during 5-8 years prior to ART had been characterized in a recent study by Zanini et al.\cite{zanini_population_2016} by whole genome deep-sequencing. The patients were selected to later have gone on to many years of fully suppressive ART. We now sequenced HIV-1 DNA from peripheral blood mononuclear cells (PBMCs) and compared these reservoir sequences to replicating HIV populations prior to ART. The timing of all available samples relative to start of treatment is summarized in \FIG{study_design}.

We found that HIV-1 DNA populations remained genetically stable for up to 18 years after start of suppressive ART, which provides evidence against viral evolution and replication as a mechanisms to maintain HIV-1 reservoirs. Furthermore, we found that variants replicating shortly before start of therapy were overrepresented in the HIV-1 DNA reservoirs indicating that proviral HIV-1 variants were turning over as long as the patients were untreated.

\section*{Results}

\subsection*{Patients and samples}
The study included 10 HIV-1 infected patients who were diagnosed in Sweden between 1990 and 2003. The patients were selected on the following criteria: 1) A relatively well-defined time of infection; 2) Treatment-naive for a minimum of 5 years; and 3) Thereafter gone on to suppressive ART (plasma HIV-1 RNA levels continuously \textless 50 copies/ml) for a minimum of 2 years. In a recent study we performed whole-genome deep sequencing of replicating HIV-1 RNA populations in 9 of the 10 patients covering the time period before they started ART (6-12 longitudinal plasma samples per patient spanning 5-8 years) \cite{zanini_population_2016}. Here we included plasma RNA sequences from the tenth patient.
Patient characteristics are summarized in \FIG{study_design} and Table 1.

\begin{table*}[tb]
\begin{tabular}{|lcccc||c|cc||c|c|}
\hline
{\bf Patient} & {\bf Gender} &{\bf Transmission} & {\bf Subtype} & {\bf Age$^{a}$} &
\multicolumn{3}{c||}{\bf HIV RNA from plasma}&
\multicolumn{2}{c|}{\bf HIV DNA from PBMCs}\\
&&&&& \#~samples&
\multicolumn{2}{|c||}{first/last since EDI$^{b}$} & time on ART$^{b}$ & \#~templates \\
 \hline
 p1 & F & HET & 01\_AE &37& 12 & 0.3 & 8.2&7.9/ 9.9/ 10.4&820/ 148/ 38 \\
 p2 & M & MSM &B       &32& 6  & 0.2  & 5.5&6.9&75\\
 p3 & M & MSM &B       &52& 10 & 0.4 & 8.4&4.6/6.7/ 7.2&243/ 102/ 108 \\
 p5 & M & MSM &B       &38& 7  & 0.4 & 5.9&4.0/ 6.3&180/ 72 \\
 p6 & M & HET &C       &31& 7  & 0.2  & 7.0&3.0/ 5.0/ 5.5&115 /15/ nd \\
 p7 & M & MSM &B       &31& 11 & 6.3$^{c}$& 16.1&6.3/ 8.4/ 8.8&88/ 279/ 108 \\
 p8 & M & MSM &B       &35& 7  & 0.2  & 6.0&8.4/ 10.6/ 10.9&180/ 55/ 175 \\
 p9 & M & MSM &B       &32& 8  & 0.3 & 8.1&7.7/ 9.7/ 10.2&60/ 72/ 72 \\
p10 & M & MSM &B       &34& 9  & 0.1  & 6.2&16.2/ 18.3/ 18.6&249/ 116/ 51 \\
p11 & M & MSM &B       &53& 7  & 0.6 & 5.6&6.4/ 8.4/ 8.8&124/ 120/ 123 \\\hline
\end{tabular}
\begin{flushleft}
\caption{{\bf Summary of patient characteristics.} $^{a}$ at diagnosis; $^{b}$ EDI: estimated date of infection; all times are given in years; $^{c}$ sequencing failed in earlier samples due to low plasma HIV-1 RNA levels.}
\end{flushleft}
\label{tab:patients}
\end{table*}

For the present study we have obtained sequence data from HIV-1 DNA in viral reservoirs by deep sequencing of the p17gag region of the HIV-1 genome in DNA prepared from PBMC. Patient characteristics are summarized in Table 1. Longitudinal PBMC samples (1-3 samples per patient spanning up to 2.6 years) were obtained obtained 3 - 18 years after start of suppressive ART (Table 1). We define viral DNA reservoirs as HIV p17gag sequences that were still present in PBMC after a minimum of 2 years of suppressive ART. The HIV-1 DNA template numbers were quantified by limiting dilution. Identical p17gag sequences were merged into haplotypes while preserving their abundance. Minor haplotypes were merged with major haplotypes if they differed by only one mutation (see Materials and Methods). Processed sequence data will be made available at \url{hiv.tuebingen.mpg.de}. Raw sequencing reads from all HIV-1 DNA samples have been deposited in the European Nucleotide archive and will be available under study accession number PRJEB13841 (sample accession numbers ERS1138001-ERS1138025).

\begin{figure*}[tb]
    \centering
    \includegraphics[width=1.7\columnwidth]{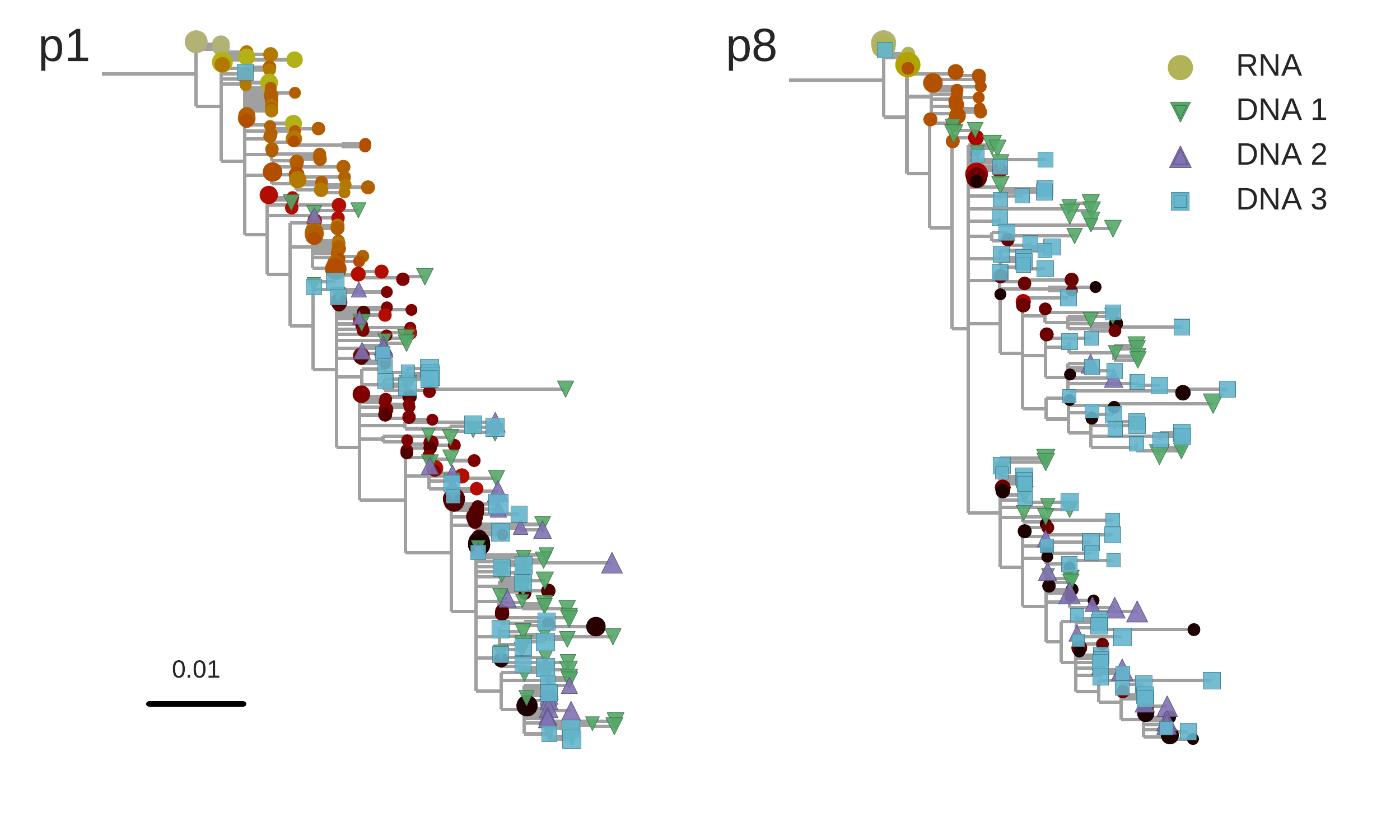}
    \caption{Reconstructed phylogenetic trees of RNA sequences (circles) and DNA (blue squares and triangles) from two patients. RNA sequences are colored by time since infection from yellow to red to black. Size of the symbols indicates the fraction of reads represented by the node. The trees were build using the software FastTree (see methods) \cite{price_fasttree_2010}.}
    \label{fig:tree_example}
\end{figure*}

\subsection*{Proviral DNA sequences reflect pretreatment RNA sequences}
The HIV-1 DNA sequences recapitulate the diversity observed in RNA sequences before treatment, often with exact sequence matches, see \FIG{tree_example} and Fig.~S3. While we observed large variations in the abundance of haplotypes with sequence read frequencies varying between 0.1\% and 50\% (see Fig.~S1), the close match between RNA and DNA sequences confirms that we characterized proviral diversity in a specific and sensitive manner.
Variation in haplotype abundance likely reflects clonal expansions \cite{josefsson_hiv-1_2013,stockenstrom_longitudinal_2015}, independent integrations of identical sequences, and resampling of the same original DNA templates during sequencing. The exact contribution by these distinct mechanisms is difficult dissect in our sequence data.

The estimated number of HIV DNA templates, the number of distinct haplotypes observed, and the fraction of haplotypes seen in multiple samples are given in Table S1. We typically recapture one third (median 0.29) of haplotypes observed at a frequency above 1\% in another sample from the same patient.

\subsection*{Hypermutated sequences are frequent in HIV-1 reservoirs}
We found that a substantial proportion (median 14\%; range 0-42\%) of the p17gag DNA sequences from the viral reservoirs were hypermutated and therefore replication incompetent (see Fig.~S2), which is consistent with other reports \cite{josefsson_hiv-1_2013,stockenstrom_longitudinal_2015,bruner_towards_2015,kieffer_g-->hypermutation_2005}. A small proportion of sequences had stop codons not obviously due to G-to-A hypermutation (average 3\%, range 0-12\%) . It is likely that a proportion of sequences without overt inactivating mutations were also replication incompetent due to mutations or deletions outside of p17gag.

Hypermutation in the HIV-1 DNA sequences complicates comparison with non-defective DNA and RNA sequences. For this reason we excluded hypermutated sequences from the main analyses, but we also performed complementary analyses that included hypermutated sequences.

\subsection*{Lack of evidence of persistent replication in HIV-1 DNA reservoirs}
It remains controversial whether or not HIV-1 reservoirs are maintained by persistent replication \cite{stockenstrom_longitudinal_2015,chomont_maintenance_2011,chomont_hiv_2009,buzon_hiv-1_2010,yukl_effect_2010,lorenzo-redondo_persistent_2016,evering_absence_2012}. We used the p17gag DNA sequences from viral reservoirs to search for evidence of sequence evolution, which would be expected to take place if the virus was replicating. The p17gag DNA sequences from viral reservoirs were obtained from 3.0 to 18.6 years after start of suppressive ART. HIV-1 RNA sequences from plasma samples obtained before start of therapy were used as reference materials.

\begin{figure*}[tb]
    \centering
    \includegraphics[width=1.5\columnwidth]{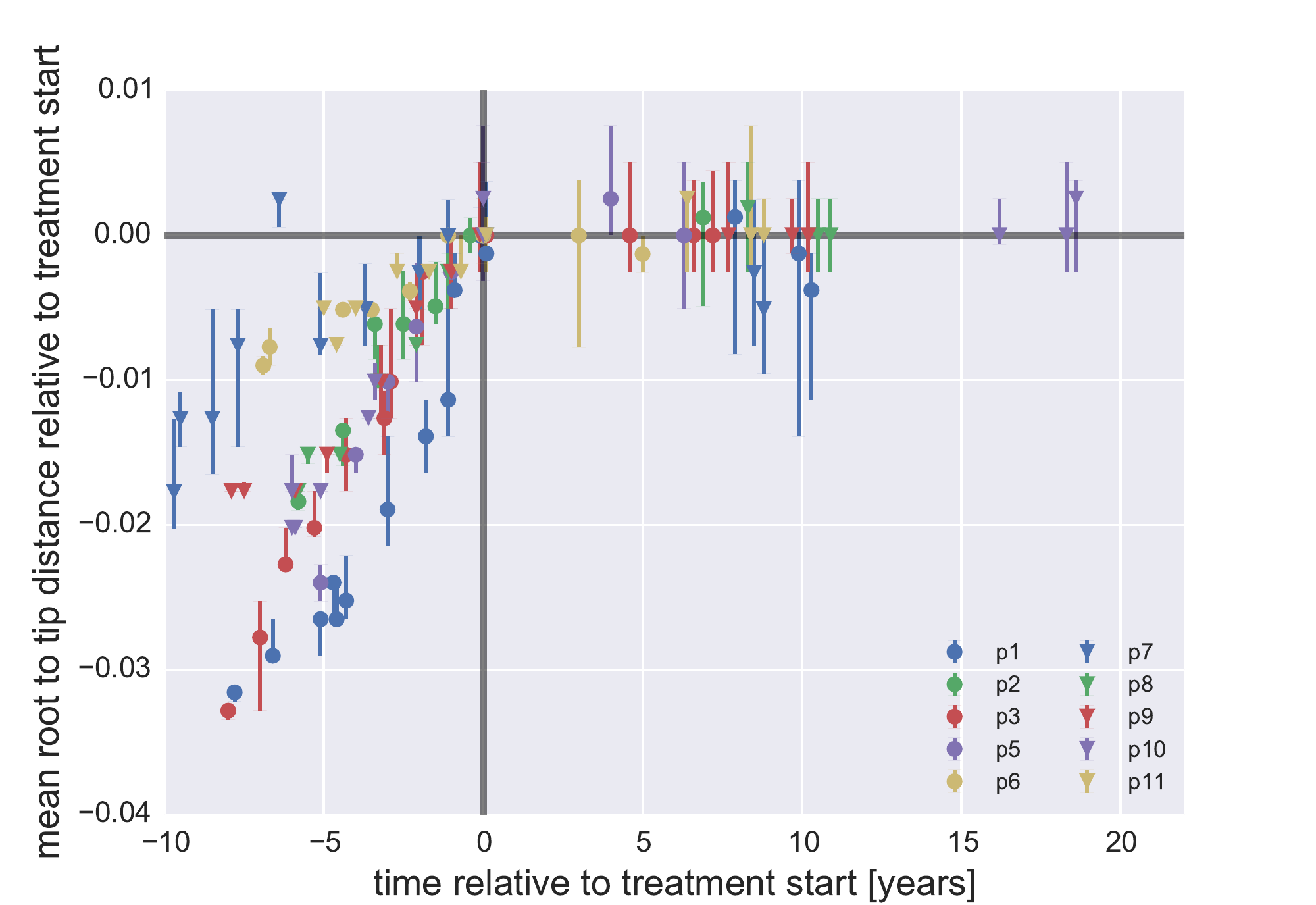}
    \caption{{\bf Root-to-tip distances.} Plasma HIV-1 RNA sequences evolve steadily before start of ART, while no evolution is observed in PBMC HIV-1 DNA sequences obtained after start of ART. The figure contains data on DNA sequence not classified as hypermutants, the analogous figure for hypermutants in shown in Fig.~S4. The error bars indicated the inter-quartile range of the root to tip distance.}
    \label{fig:root_to_tip}
\end{figure*}

Root-to-tip distances for plasma RNA populations and PBMC DNA populations were calculated relative to the major RNA haplotype in the first plasma sample. \FIG{root_to_tip} shows temporal changes of root-to-tip distances in HIV-1 RNA and DNA populations obtained before and after start of suppressive ART, respectively. As previously shown, plasma HIV-1 RNA populations obtained before start of ART evolved at a relatively constant rate \cite{zanini_population_2016}, as evidenced by a steady increase of average root-to-tip distances over time. In sharp contrast, HIV-1 DNA populations obtained after 3 - 18 years of suppressive therapy showed stable root-to-tip distances. Hypermutated DNA sequences showed larger root-to-tip distances, but also these distances were stable over time (Fig.~S4)

Table 2 shows the rate of evolution before and after start of suppressive ART. Before start of therapy we observed statistically significant evolution of plasma RNA sequences with rates $1$ to $4\times 10^{-3}$/year in all 10 patients. In contrast, no statistically significant evolution was observed in DNA reservoirs during up to 18 years of suppressive ART.

Collectively, our results do not provide support for persistent HIV-1 replication as a mechanism to maintain the HIV-1 reservoir during suppressive therapy.

\begin{table}[tb]
    \centering
    \begin{tabular}{|c|r|r||r|r|}
    \hline
{\bf patient} & \multicolumn{2}{c||}{{\bf RNA rate}} & \multicolumn{2}{c|}{{\bf DNA rate}}\\
& $[\mathrm{year}^{-1}]$& $p$-value & $[\mathrm{year}^{-1}]$& $p$-value \\\hline
p1  & $4.2\times 10^{-3}$ & $<10^{-6}$ &  $1\times 10^{-4}$ &  0.866 \\
p2  & $3.3\times 10^{-3}$ & $<10^{-3}$ &  $2\times 10^{-4}$     & -- \\
p3  & $4.2\times 10^{-3}$ & $<10^{-6}$ &  $0$                   &1.0 \\
p5  & $4.6\times 10^{-3}$ & $<10^{-3}$ &  $1\times 10^{-4}$     & 0.93 \\
p6  & $1.3\times 10^{-3}$ & $<10^{-3}$ &  $-6\times 10^{-4}$    &  0.67 \\
p7  & $1.4\times 10^{-3}$ & $<10^{-2}$ &  $-8\times 10^{-4}$    &  0.3 \\
p8  & $3.0\times 10^{-3}$ & $<10^{-4}$ &  $2\times 10^{-5}$     &  0.94 \\
p9  & $2.4\times 10^{-3}$ & $<10^{-4}$ &  $3\times 10^{-7}$     &  0.87 \\
p10 & $3.5\times 10^{-3}$ & $<10^{-5}$ &  $-1\times 10^{-4}$    &  0.55 \\
p11 & $1.1\times 10^{-3}$ & $<10^{-2}$ &  $6\times 10^{-5}$     & 0.6 \\\hline
    \end{tabular}
    \begin{flushleft}
    \caption{Rates of evolution in plasma RNA and PBMC DNA sequences obtained before start and after start of suppressive ART, respectively. }
    \end{flushleft}
    \label{tab:rates}
\end{table}

\subsection*{Time for deposition of reservoir DNA sequences}
\begin{figure}[htb]
    \centering
    \includegraphics[width=\columnwidth]{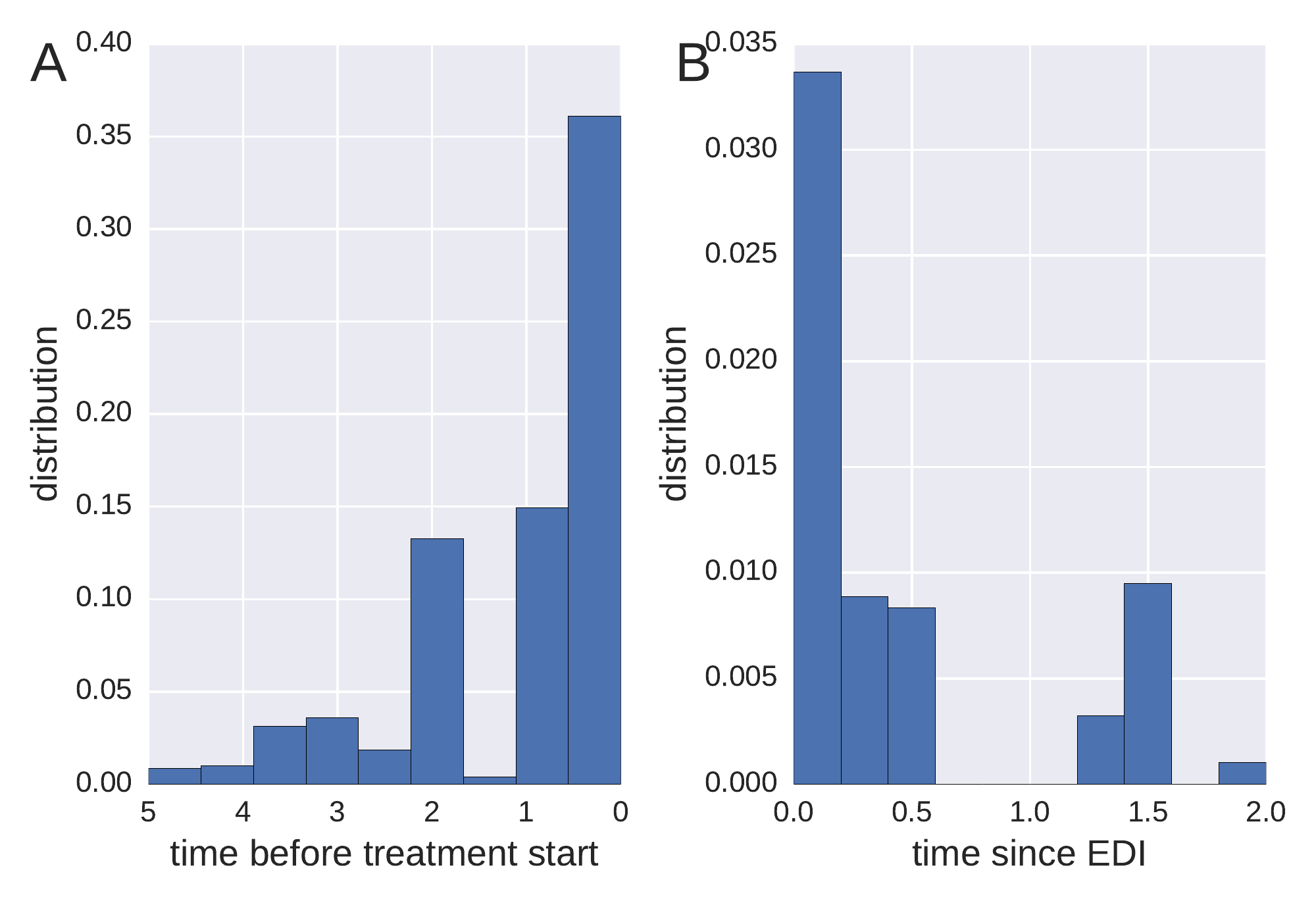}
\caption{{\bf Probable origin of sequences in the DNA reservoir.} Panel A shows the time difference between the start of treatment and the RNA sample that is most likely to have generated at sequence from the reservoir. A large proportion of the reservoir DNA sequences are most closely related to RNA sequences in the plasma samples obtained during the last year before start of therapy. Panel B focuses on DNA sequences resembling RNA sequences obtained during the first two years after estimated time of infection. These sequences represent a small fraction of all reads obtained and are observed in 50\% of patients only.}
    \label{fig:seeding}
\end{figure}

To investigate when PBMC HIV-1 DNA variants had been deposited the viral reservoirs we compared the on-treatment PBMC DNA sequences with the longitudinal pre-treatment plasma RNA sequences (\FIG{seeding} and Fig.~S5).
For each p17gag DNA sequence, we determined the RNA sample and haplotype that was the most likely source.
By this procedure, most proviral sequences were assigned to the plasma samples closest to the start of treatment (\FIG{seeding}, panel A).
The representation of RNA haplotypes from earlier plasma samples dropped rapidly over 1-2 years.
However, sequences representing earlier plasma sampling time points were also found as minor variants among the p17gag DNA sequences. Among these minor variants, DNA sequences matching the variant dominating the first plasma samples obtained within 6 month after infection were overrepresented in some patients (\FIG{seeding}, panel B). Thus, the proportion of DNA sequences matching the initial plasma HIV-1 RNA haplotype was 14\%, 2.4\%, 42\%, \textless 1\%, and 6.9\% in patients 2, 3, 6, 8 and 11, respectively.

The overrepresentation in the HIV-1 DNA reservoir of viral variants that replicated shortly before start of suppressive ART indicates that cells carrying defective and non-defective proviral variants were turning over as long as the patients were untreated and that suppressive ART halted this turnover. The width of the peak in \FIG{seeding}A suggest a half-life on the order of one year.

\section*{Discussion}

In this study we have investigated the composition and turnover of HIV-1 DNA sequences in viral reservoirs in patients on long-term suppressive therapy. Reservoir HIV-1 DNA populations were remarkably stable and showed no signs of ongoing replication. We also traced when during the course of HIV infection viruses in the the DNA reservoirs had been deposited and found that they mainly derived from the last year(s) before start of suppressive therapy.

Our study provides evidence against persistent HIV-1 replication as a mechanism for maintenance of HIV-1 reservoirs during suppressive therapy. This is at variance with a recent report by Lorenzo-Redondo et al.~\cite{lorenzo-redondo_persistent_2016} and a few earlier reports
\cite{yukl_effect_2010,buzon_hiv-1_2010}, but agrees with several other earlier studies \cite{stockenstrom_longitudinal_2015,besson_hiv-1_2014,josefsson_hiv-1_2013,kearney_lack_2014,gunthard_evolution_1999,evering_absence_2012,kieffer_absence_2004}. Lorenzo-Redondo et al.~\cite{lorenzo-redondo_persistent_2016} compared genetic diversity in samples HIV-1 RNA in plasma at start of therapy with HIV-1 DNA sequences obtained from blood and tissues at baseline, three and six months after the start of treatment. They report a signal of evolution between the different time points at an extraordinary high rate ($7.4-12\times 10^{-3}$ changes per site per year) -- about 5-fold higher than typically observed in gag and pol of replicating RNA populations.
This observation is incompatible with the lack of observable changes in reservoir sequences over 20 times longer time intervals reported here.
Without longitudinal data on the evolution of the HIV-1 population prior to treatment, the nature of the change reported by Lorenzo-Redondo et al.~\cite{lorenzo-redondo_persistent_2016} is difficult to discern. One possible explanation for the apparent conflict between the two studies could be that after onset of therapy, short-lived cells that sampled the most recent circulating virus populations start to disappear, leaving longer-lived cells that sample deeper into the history of the infection. This scenario would not correspond to evolution, but quite oppositely a sampling of earlier variants. Another difference is that Lorenzo-Redondo et al.~\cite{lorenzo-redondo_persistent_2016} investigated HIV-1 DNA sequences in tissue as well as PBMC samples whereas we only studied PBMC samples. However, tissue and blood HIV-1 DNA variants should be well-mixed over the time frame that we investigate \cite{josefsson_hiv-1_2013,stockenstrom_longitudinal_2015,lorenzo-redondo_persistent_2016}. Both Lorenzo-Redondo et al.~\cite{lorenzo-redondo_persistent_2016} and we studied DNA sequences in HIV-1 reservoirs, which are known to contain a high proportion of defective virus. These proviruses serve as markers of T-cell clones, rather than replication-competent virus.
Hence the absence of evolution or turnover of provirus that we found does not exclude the possibility that there is replication and evolution of replication-competent virus in reservoirs.
However, if such replication exists, it happens very low levels that do not contribute substantially to the pool of proviral DNA in PBMCs.
To enrich for replication competent and putatively evolving virus, QVOA followed by sequencing of virus released into to supernatants should performed, rather than sequencing of total HIV-1 DNA as done by Lorenzo-Redondo et al.~\cite{lorenzo-redondo_persistent_2016}, us and others \cite{josefsson_hiv-1_2013,kieffer_g-->hypermutation_2005,evering_absence_2012}.
In agreement with our finding of genetic stability in the DNA reservoirs Josefsson et al.~\cite{josefsson_hiv-1_2013} and Stockenstrom et al.~\cite{stockenstrom_longitudinal_2015} have reported that defective HIV-1 DNA integrants present during long-term effective ART appear to be maintained by proliferation and longevity of infected cells rather than by ongoing viral replication.

Because we had access to detailed longitudinal data on the evolution of the plasma HIV-1 RNA population from time of infection to start of suppressive ART, we could trace when during the course of untreated HIV-1 infection the viruses in the DNA reservoirs had been deposited. We found that a majority of variants in the HIV-1 DNA reservoirs were derived from HIV-1 RNA variants that had actively replicated during the last year(s) before start of suppressive ART, with no evidence for evolution after treatment start. Frenkel et al.\cite{frenkel_multiple_2003}, in contrast to us, reported persistence of a greater number of early compared to recent viruses in a few children on suppressive ART; more research is warranted to assess the origin of this difference.

Defective HIV-1 proviruses can be regarded as unique in vivo labels of individual memory CD4 cell clones which can be used to track their fate similar to sequencing of T-cell receptors \cite{robins_immunosequencing:_2013}. This strategy was used by  Imamichi et al.\cite{imamichi_lifespan_2014} to demonstrate that a T-cell clone persisted more than 17 years. Similarly, prenatally formed T-cell receptors shared by twins have been reported to have lifetimes $>30$ years \cite{pogorelyy_persisting_2016}.
During suppressive ART the turnover of infected memory CD4 cell clones is likely to follow the same dynamics as in uninfected people.
In contrast, we observe a strong overrepresentation in the reservoirs of ``late'' HIV-1 RNA variants, which indicates that HIV-1 target cells, primarily CD4+ T-lymphocytes, were turning over with a half-life of about one year in absence of treatment. This turnover was dramatically slowed by suppressive ART. Earlier studies, based on different types of labelling of CD4 cells, have indicated a 3 - 4 fold increased rate of CD4 cell death in untreated HIV-1-infected patients as compared with uninfected persons and patients on suppressive ART \cite{hellerstein_directly_1999,mccune_factors_2000,ribeiro_vivo_2002}.
The more dramatic difference we observe is likely explained by different methodologies. Earlier studies estimated the lifespan of individual cells whereas we primarily have estimated the lifespan of CD4 cell clones carrying defective proviruses (i.e.~infected cells as well as their daughter cells).

Our study has several limitations. We have not sorted cells and therefore cannot investigate if there are differences in HIV-1 turnover between different types and subsets of cells, such as memory CD4 cells and their subsets. However, it is reasonable to assume that a majority of our HIV-1 DNA sequences came from memory CD4 cells because others have shown that these cells constitute the main HIV-1 reservoir  \cite{eriksson_comparative_2013,chun_quantification_1997,chun_vivo_1995}.
We sequenced a relatively short region of the HIV-1 genome and therefore cannot reliably distinguish between replication-competent and defective viruses.
While we observe no evolution in these proviral DNA sequences, we cannot rule out the possibility that a small subset of viruses indeed was replicating but remained undetected among the many replication-incompetent viruses.
We observed large variations in the  abundance  of  sequence haplotypes that likely reflect both clonal expansions  \cite{josefsson_hiv-1_2013,stockenstrom_longitudinal_2015}, independent integrations of identical sequences, and resampling of the same original DNA templates during sequencing.
With our sequencing method we could not exactly determine the relative contribution by these distinct mechanisms. We are attempting Primer ID sequencing \cite{jabara_accurate_2011} to even better understand the in vivo dynamics of different viral haplotypes.

In summary, we provide compelling evidence against persistent viral replication as a mechanism to maintain the latent HIV-1 DNA reservoir during suppressive therapy. Furthermore, we show that most latently infected cells during long-term suppressive ART are infected shortly before ART start and that the rate of T-cell turnover is reduced upon starting suppressive ART.

\section*{Materials and methods}
\paragraph*{Ethical statement}
The study was conducted according to the Declaration of Helsinki. Ethical approval was granted by the Regional Ethical Review board in Stockholm, Sweden (Dnr 2012/505 and 2014/646). Patients participating in the study gave written and oral informed consent to participate.

\paragraph*{Patients}
The study included 10 HIV-1-infected patients who were diagnosed in Sweden between 1990 and 2003. Prior to the present study the patients were included in a recent study on the population genomics of intrapatient HIV-1 evolution \cite{zanini_population_2016}. The patients were selected based on the following inclusion criteria: 1) A relatively well-defined time of infection based on a negative HIV antibody test less than two years before a first positive test or a laboratory documented primary HIV infection; 2) No ART during a minimum of approximately five years following diagnosis; 3) Availability of biobank plasma samples covering this time period; and 4) Later have started successful ART (plasma viral levels $<50$ copies/µl) for a minimum of two years. As previously described 6 - 12 plasma samples per patient were retrieved from biobanks and used for full-genome HIV-1 RNA sequencing \cite{zanini_population_2016}. The same patient nomenclature is used in both studies. For the present study the same patient were asked to donate 70 ml of fresh EDTA-treated blood on up to three occasions over a time period of 2.5 years. These blood samples were obtained 3 - 18 years after start of successful ART. Estimated time of infection (ETI) was calculated as previously described using clinical and laboratory findings including Fiebig staging and BED testing \cite{zanini_population_2016}. Information about the patients and the samples are summarized in Table 1.

\paragraph*{HIV-1 RNA sequencing from plasma}
Whole-genome deep-sequencing of virus RNA populations in plasma samples obtained before start of therapy was performed as previously described \cite{zanini_population_2016}. In short, total RNA in plasma was extracted using RNeasy® Lipid Tissue Mini Kit (Qiagen Cat No. 74804) and amplified using  a one-step RT-PCR with outer primers for six overlapping regions and Superscript ® III One-Step RT-PCR  with Platinum ® Taq High Fidelity High Enzyme Mix (Invitrogen, Carlsbad, California, US). An optimized Illumina Nextera XT library preparation protocol was used together with a kit from the same supplier to build DNA libraries, which were sequenced on the Illumina MiSeq instrument with 2 x 250bp or 2x 300bp sequencing kits (MS-102-2003/MS-10-3003). For the present study a part of the p17gag region of the HIV-1 genome (see below) was extracted from the entire full-genome RNA data set. The median number of high quality reads covering the entire p17 sequence was 146 (inter-quartile range 56 - 400) and the cDNA template numbers are available in Zanini et al.\cite{zanini_population_2016}.

\paragraph*{HIV-1 DNA sequencing from PBMCs}

Approximately 70 ml of fresh whole blood was obtained in 7 Vacutainer tubes with EDTA as anticoagulant. PMBC were isolated by Ficoll-Paque PLUS (GE Healthcare Bio-Sciences AB, Uppsala, Sweden) centrifugation according to the instructions by the manufacturer. Total DNA was extracted from PBMC using the OMEGA E.Z.N.A® Blood DNA Mini Kit (Omega bio-tek, Norcross, Georgia) or the QIAamp DNA Blood Mini Kit (Qiagen GmbH, Hilden, Germany) according to the instructions by the manufacturer.
The amount of DNA was measured with Qubit® dsDNA HS Assay Kit (Invitrogen™ Eugene, Oregon, USA).
Patient-specific nested primers (Integrated DNA Technologies) were used to amplify a 387-bp long portion of the p17gag gene corresponding to positions 787 to 1173 in the HxB2 reference sequence. The primers were designed based on the plasma RNA sequences from each patient (Tab.~S2).
Outer primers were used together with Platinum® Taq DNA Polymerase High Fidelity (Invitrogen™ Carlsbad, California, US) for the first PCR. The program started with a denaturation step at 94°C for 2 min followed by 15 PCR cycles of denaturation at 94°C for 20 s, annealing at 50°C for 20 s and extension at 72°C for 30 s and a final extension step at 72°C for 6 min. For the second PCR, 2.5 μl of the product from the first PCR was amplified with inner primers and the cycle profile and enzyme as for the first PCR. Amplified DNA was purified using Agencourt AMPure XP (Beckman Coulter Beverly, Massachusetts) and quantified using Qubit.
For each sample the number of HIV-1 DNA templates used for sequencing was roughly quantified in triplicate by limiting dilution using the same PCR conditions, three dilutions (usually 0.5, 0.1, 0.02 µg of DNA) and Poisson statistics.
Control experiments were performed to evaluate PCR-induced recombination using the plasmids NL4-3 and SF162, which were spiked in equal proportion into human DNA and amplified using the same PCR conditions as above. The results showed that there was minimal PCR-induced recombination in this short amplicon.

\paragraph*{Sequencing and read processing}
The HIV specific primers were flanked by NexteraXT adapters. To construct sequencing libraries, indices and sequencing primers were added in 12-15 cycles of additional PCR.
Amplicons were sequenced on an Illumina MiSeq machine with 2x250 cycle kits. Between 6,500 and 190,000 (median 35,000) paired-end reads were generated per sample. The overlapping paired-end sequencing reads were merged to create synthetic reads spanning the entire p17 amplicon. In case of disagreement between paired reads, the nucleotide on the read with the higher quality score was used. We counted the number of times a particular p17 sequence was observed and did subsequent analysis with read-abundance pairs.
To reduce the influence of sequencing and PCR errors, we combined rare sequences (below frequency 0.002) with common sequences if they differed by no more than one position. Specifically, starting with the rarest sequences, we merged rare sequences with the most common sequence that was one base away. The cutoff 0.002 is the typical error frequency of the pipeline as determined earlier \cite{zanini_population_2016}.
All analysis is done in Python using the libraries numpy, biopython, and matplotlib  \cite{cock_biopython:_2009,van_der_walt_numpy_2011,hunter_matplotlib:_2007}.

\paragraph*{Hypermutation detection}
To classify sequences into obvious hypermutants and sequences representative of circulating virus, we counted mutations at positions that are not variable in the RNA samples obtained prior to therapy. If more than 4 mutations were observed and at least half of them were \texttt{G}$\to$\texttt{A}, the sequence was considered a hypermutant. The distribution of the different transition mutations relative to the closest genome found in RNA samples are shown in Fig.~S2 for reads classified as hypermutants or not. Results we obtained for sequences classified as non-hypermutants are very similar to results obtained when using only sequences without stop codons.

\paragraph*{Phylogenetic analysis}
We reconstructed phylogenetic trees using the approximate maximum likelihood method implemented by FastTree \cite{price_fasttree_2010}. Tips were annotated with frequency, source and sample date using custom python scripts.

\paragraph*{Statistical analysis}
Root-to-tip distances were calculated as the average distance between a sample and the founder sequence approximated by the consensus sequence of the first RNA sample. To determine the rate of evolution in absence of treatment, this root-to-tip distance was regressed against time. To determine the rate of evolution on treatment, the root-to-tip sequence of the last RNA sample and the DNA samples was regressed against time. To determine the most likely seeding time for a p17gag DNA sequence obtained from PBMCs, we calculated the likelihood of sampling this sequence given the SNP frequencies in each RNA sample and assigned the sequence to the sample where this likelihood was highest.

\paragraph*{Acknowledgements}
This work was supported by the European Research Council through grant Stg.~260686 and the Swedish Research Council trough grant K2014-57X-09935. We would also like to express our gratitude to the study participants.

\bibliography{bibliography}

\clearpage
\section*{Supplementary Figures and Tables}
\onecolumngrid
\renewcommand\thefigure{S\arabic{figure}}
\setcounter{figure}{0}
\renewcommand\thetable{S\arabic{table}}
\setcounter{table}{0}

\begin{figure}[hb]
    \centering
    \includegraphics[width=0.7\columnwidth]{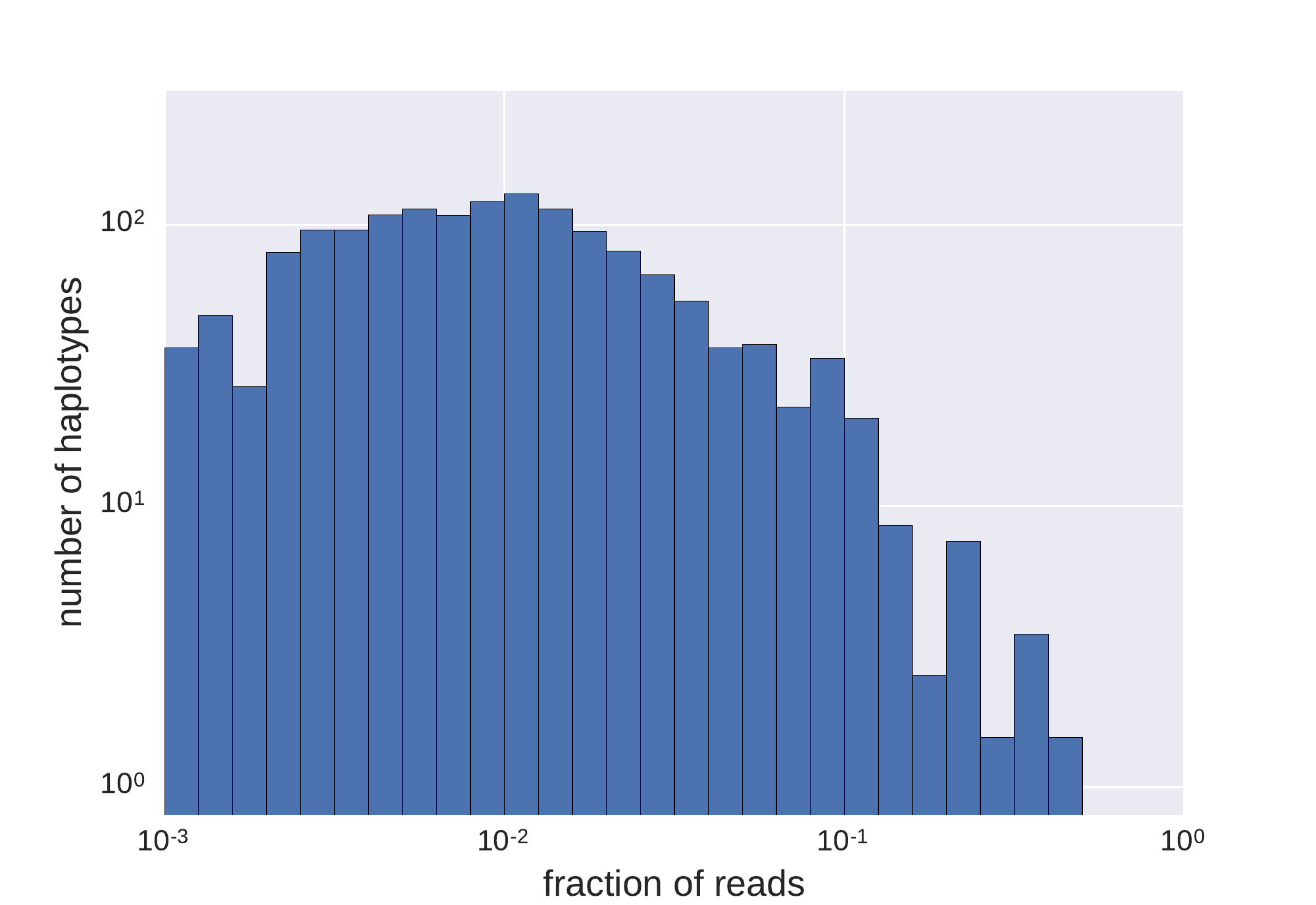}
    \caption{Distributions of frequencies of haplotypes. We observe a wide variation of haplotype frequencies -- measured by fraction of reads -- from below 0.001 to above 0.5. The majority of haplotypes are seen at frequencies around 0.01. Note that both scales are logarithmic. }
    \label{fig:haplo_frequencies}
\end{figure}

\begin{figure}[hb]
    \centering
    \includegraphics[width=0.7\columnwidth]{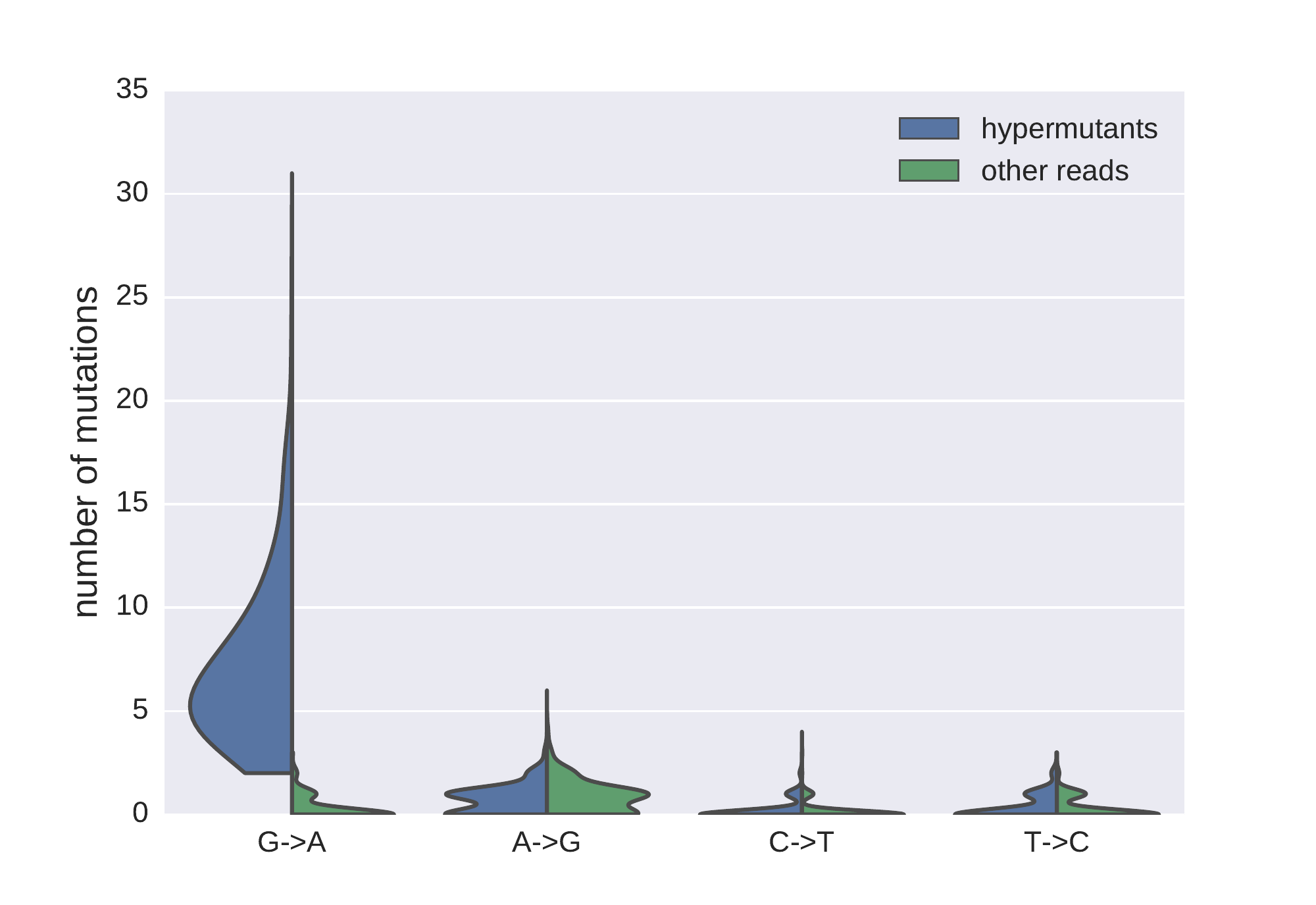}
    \caption{Distributions of mutations classified as hypermutants or regular reads.}
    \label{fig:hypermuts}
\end{figure}

\begin{figure*}
    \centering
    \includegraphics[width=1.3\columnwidth,angle=90]{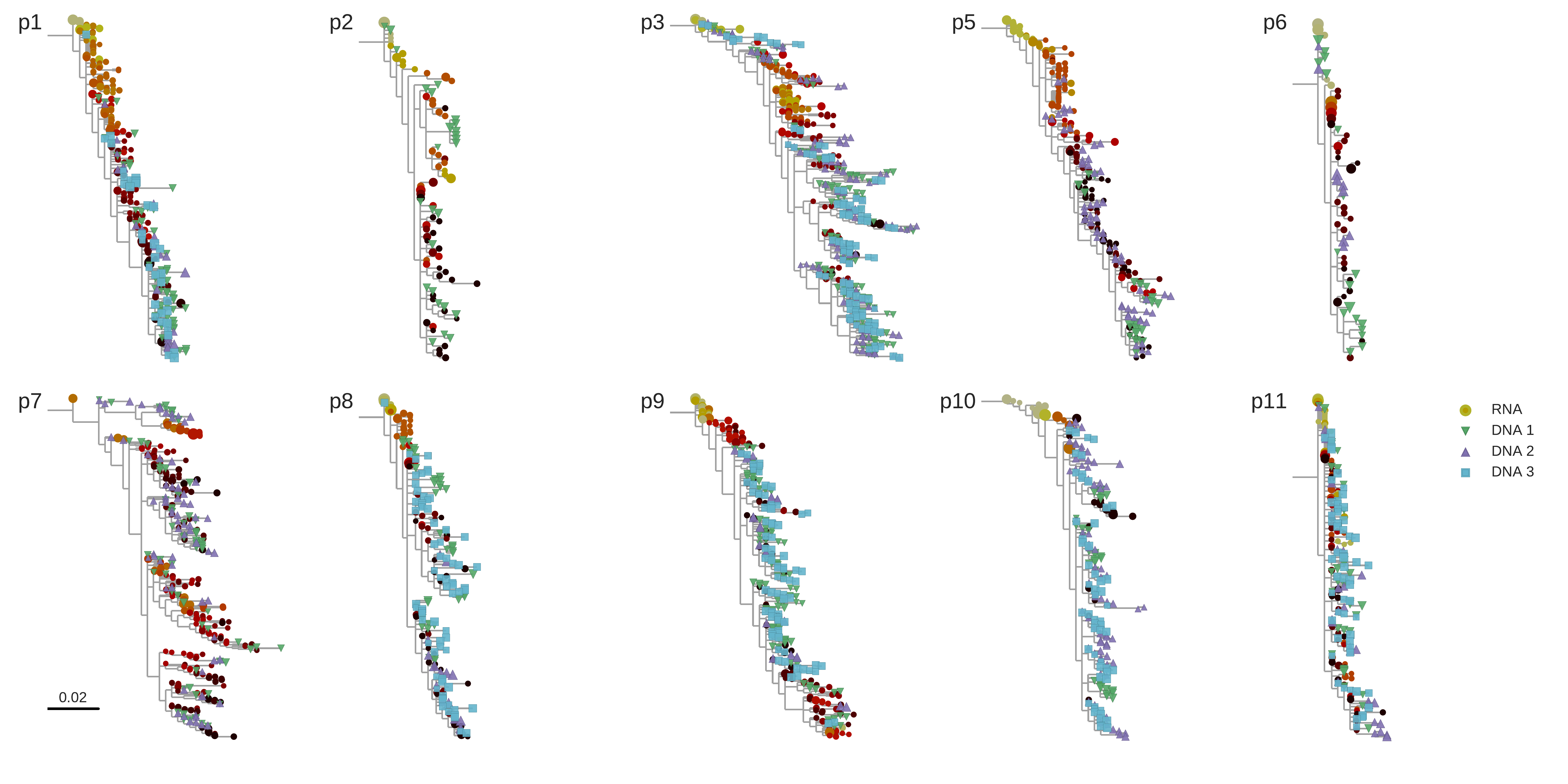}
    \caption{Phylogenetic trees of RNA and DNA samples.}
    \label{fig:supp_trees}
\end{figure*}
\clearpage
\begin{figure*}
    \centering
    \includegraphics[width=0.7\columnwidth]{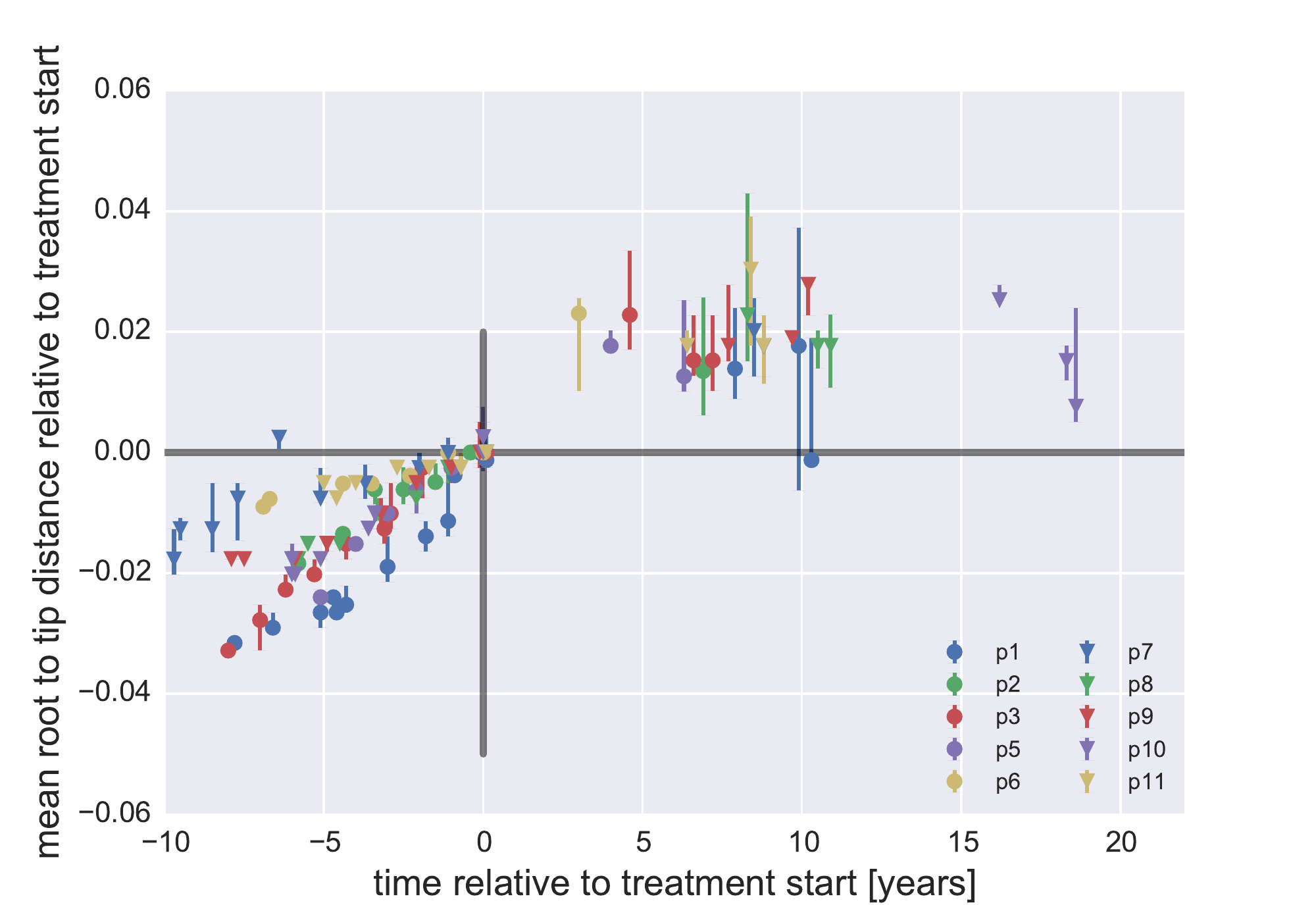}
    \caption{Average root-to-tip distances for plasma HIV-1 RNA sequences obtained before start of ART and PBMC HIV-1 DNA sequences obtained after start of ART. This figure is analogous to Fig.~2 in the main text, but shows root-to-tip distance of DNA sequences classified as hypermutants. While hypermutant sequences are between 2 and 4\% more distance from the approximate founder sequence, the distances do not change over time.}
    \label{fig:root_to_tip_hyper}
\end{figure*}

\begin{figure*}
    \centering
    \includegraphics[width=\columnwidth]{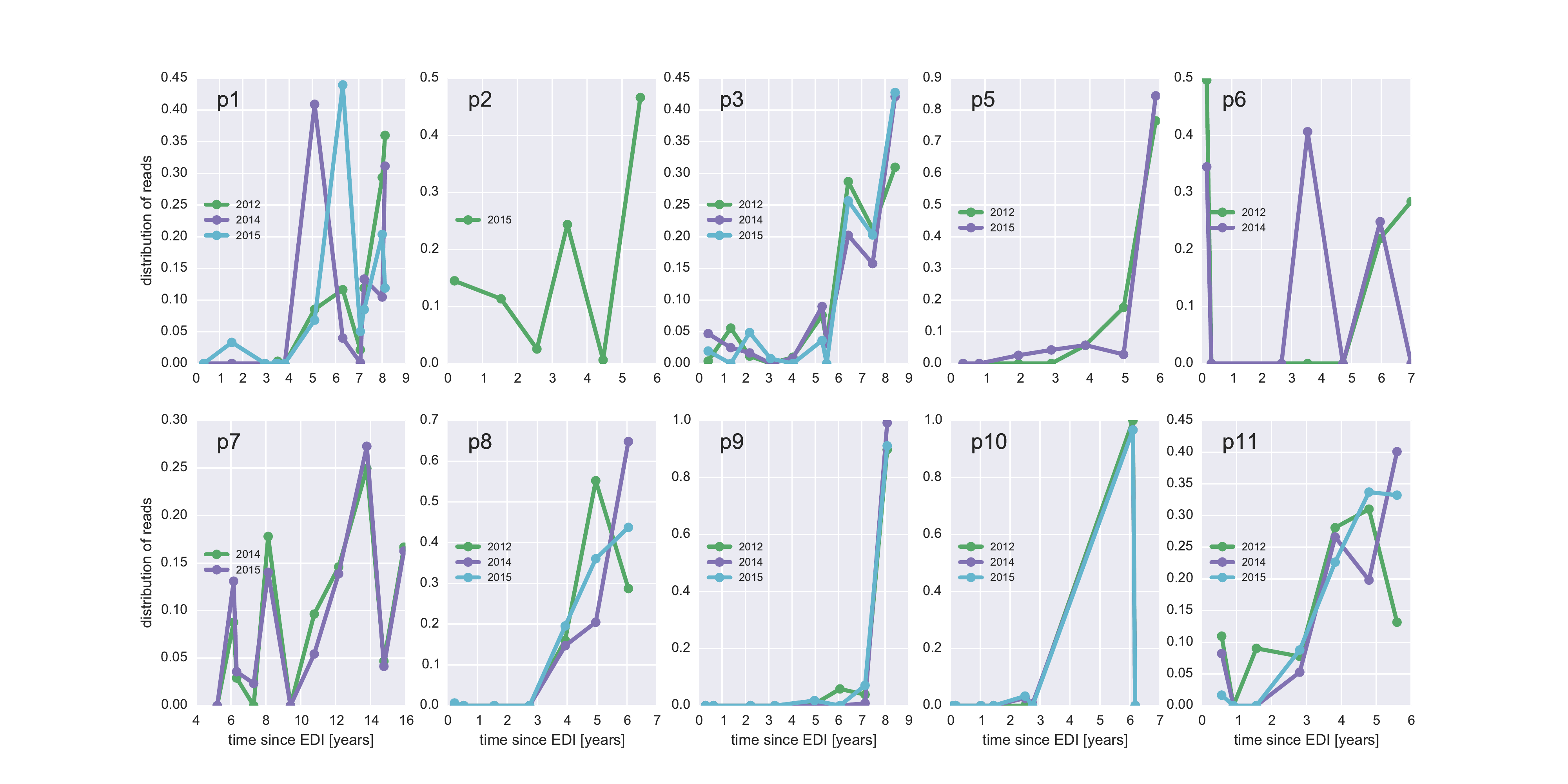}
    \caption{The distribution of plausible seeding times of reservoir sequences. For each read obtained from DNA samples, the RNA sample from which this read is most likely drawn is determined. The plots show the distribution of these most likely origin samples across all available RNA samples.}
    \label{fig:supp_timeline}
\end{figure*}

\clearpage

\begin{sidewaystable}[htb]
    \scriptsize
    \centering
    \begin{tabular}{|cccc|ccc|ccc|cc|}\hline
patient &   sample &    \#reads &   \#templates &   \#seq$>0.002$   &   \#seq$>0.01$    &   \% recaptured & \% hyper    & \#stop good&  \#stop hyper&   \% stop good&   \%stop hyper\\ \hline
p1  &   2012-10-02  &   19363   &   820 &   58  &   27  &   33 &    13  &   8   &   1.85    &   1.4 &   83\\
p1  &   2014-10-07  &   36454   &   148 &   23  &   13  &   31 &    14  &   9   &   0.81    &   1.3 &   36\\
p1  &   2015-02-12  &   44434   &   38  &   36  &   22  &   36 &    17  &   7   &   1.43    &   0.8 &   85\\
p2  &   2015-06-09  &   16586   &   75  &   26  &   12  &   NA &    29  &   9   &   1.65    &   3.9 &   99\\
p3  &   2012-09-18  &   23347   &   243 &   116 &   26  &   50 &    19  &   8   &   1.23    &   4.6 &   75\\
p3  &   2014-10-13  &   28450   &   102 &   117 &   31  &   45 &    30  &   12  &   1.78    &   4.9 &   82\\
p3  &   2015-04-24  &   44850   &   108 &   98  &   34  &   38 &    24  &   10  &   1.21    &   5.0 &   61\\
p5  &   2012-10-26  &   11190   &   180 &   23  &   14  &   14 &    21  &   7   &   1.99    &   0.9 &   94\\
p5  &   2015-03-16  &   39762   &   72  &   56  &   29  &   7 &     6   &   8   &   1.36    &   3.8 &   99\\
p6  &   2012-10-24  &   191468  &   115 &   17  &   11  &   9 &     10  &   9   &   0.94    &   1.0 &   60\\
p6  &   2014-11-03  &   33887   &   15  &   9   &   3   &   33 &    0   &   9   &   nan     &   1.0 &   nan\\
p7  &   2014-12-01  &   35565   &   28  &   71  &   43  &   0   &   13  &   5   &   1.43    &   2.5 &   88\\
p7  &   2015-04-17  &   38323   &   108 &   67  &   41  &   0   &   11  &   5   &   1.34    &   0.5 &   94\\
p8  &   2012-09-21  &   6553    &   180 &   32  &   12  &   17  &   42  &   6   &   1.75    &   1.5 &   87\\
p8  &   2014-11-18  &   75473   &   279 &   16  &   11  &   0   &   30  &   6   &   2.00    &   0.7 &   99\\
p8  &   2015-04-07  &   18750   &   175 &   68  &   35  &   6   &   33  &   6   &   1.44    &   3.6 &   88\\
p9  &   2012-10-05  &   43306   &   60  &   64  &   43  &   19 &    13  &   5   &   0.52    &   1.1 &   44\\
p9  &   2014-10-02  &   11620   &   72  &   27  &   16  &   38 &    0   &   3   &   0.0     &   0.5 &   100\\
p9  &   2015-03-18  &   65003   &   72  &   71  &   38  &   16 &    12  &   6   &   1.86    &   0.7 &   98\\
p10 &   2012-10-09  &   7513    &   249 &   32  &   16  &   25 &    4   &   02  &   2.0     &   0.3 &   100\\
p10 &   2014-10-24  &   50537   &   116 &   74  &   29  &   28 &    14  &   16  &   1.22    &   7.1 &   99\\
p10 &   2015-02-27  &   59041   &   51  &   55  &   29  &   41 &    00  &   6   &   0.55    &   0.6 &   27\\
p11 &   2012-10-10  &   47701   &   124 &   36  &   23  &   30 &    16  &   20  &   1.00    &   6.0 &   86\\
p11 &   2014-10-22  &   32126   &   120 &   29  &   17  &   53 &    28  &   21  &   2.08    &   8.4 &   84\\
p11 &   2015-02-25  &   6834    &   123 &   57  &   37  &   35 &    33  &   11  &   0.99    &   3.7 &   75\\\hline
\end{tabular}
\begin{flushleft}
    \normalsize
    {Sequencing and hypermutation statistic of all samples. ``good'' refers to proviral sequences that are not obviously defective, while ``hyper'' refers to those with an excess of \texttt{G}$\to$\texttt{A} mutations.}
\end{flushleft}
    \label{tab:hypermut_stats}
\end{sidewaystable}

\begin{sidewaystable}[htb]
    \scriptsize
    \centering
    \begin{tabular}{|l|l|l|}\hline
Patient/Plasmid & Primer name & Sequence \\ \hline
pAll &              $\mathtt{all^{a}\_fw^{b}{1}^{c}\_689^{d}\_705^{e}}$                            & \verb|ACG CAG GAC TCG GCT TGC | \\
pAll &              \verb|all_fw2(nex)|${}^{f}$\verb|_770_787|                          & $(^{g}$\verb|TGC TCG GCA GCG TCA GAT GTG TAT AAG AGA CAG|$)^{h}$\verb|CGG AGG CTA GAA GGA GAG| \\
p1-p5, p7, p8-p11 & \verb|p1_p2_p3_p4_p5_p7_p8_p9_p10_p11_rev1_1339_1321|  & \verb|AAT CTT GTG GGG TGG CTC C| \\
p6 &                \verb|p6_rev1_1339_1321|                               & \verb|AAT CTG CTG GRG TGG CTC C| \\
p1, p7, p8, p11 &   \verb|p1_p7_p8_p11_rev2(nex)_1101_1176|                & \verb|(GTC TCG TGG GCT CGG AGA TGT GTA TAA GAG ACA G)GT ATA GGG TAA TTT TGG CTG| \\
p2 &                \verb|p2_rev2(nex)_1191-1176|                          & \verb|(GTC TCG TGG GCT CGG AGA TGT GTA TAA GAG ACA G)GT ATG GGG TAA TTT TGG CTG| \\
p3 &                \verb|p3_rev2(nex)_1191-1176|                          & \verb|(GTC TCG TGG GCT CGG AGA TGT GTA TAA GAG ACA G)GT ATC GGG TAA TTT TGG CTG| \\
p5 &                \verb|p5 _rev2(nex)_1191-1176|                         & \verb|(GTC TCG TGG GCT CGG AGA TGT GTA TAA GAG ACA G)GT ATA GGG WAA TTT TGG CTG| \\
p6 &                \verb|p6_rev2(nex)_1191-1173|                          & \verb|(GTC TCG TGG GCT CGG AGA TGT GTA TAA GAG ACA G)GT ATA GGA TAA TTT TGG CTG| \\
p9, p10 &           \verb|p9_p10_rev2(nex)_1191-1176|                      & \verb|(GTC TCG TGG GCT CGG AGA TGT GTA TAA GAG ACA G)GT ATA GGG TAA TTT TGR CTG| \\
NL4-3 &             \verb|NL4-3_fw1_689_705|                               & \verb|ACG CAG GAC TCG GCT TGC| \\
NL4-3 &             \verb|NL4-3_rev1_1339_1321|                            & \verb|AAT CTT GTG GGG TGG CTC C| \\
NL4-3 &             \verb|NL4-3_fw2(nex)_770_787|                          & \verb|(TGC TCG GCA GCG TCA GAT GTG TAT AAG AGA CAG) CGG AGG CTA GAA GGA GAG| \\
NL4-3 &             \verb|NL4-3_rev2(nex)_1191_1173|                       & \verb|(GTC TCG TGG GCT CGG AGA TGT GTA TAA GAG ACA G) TAT AGG GTA ATT TTG GCT G| \\ \hline
    \end{tabular}
    \normalsize
    \begin{flushleft}
    {{\bf Primer table.}
$^{a}$ Patient number.
$^{b}$ Primer direction.
$^{c}$ Primer position, 1=outer primer 2=inner primer.
$^{d}$ Start position of the primer according to HXB2 coordinates.
$^{e}$ Stop position of the primer according to HXB2 coordinates.
$^{f}$ Nextera adapter.
$^{g}$ Start position of the NexteraXT adapter.
$^{h}$ Stop position of the NexteraXT adapter.
}
\end{flushleft}
\label{tab:primers}
\end{sidewaystable}

\end{document}